\documentclass[preprint,aps,prb]{revtex4}
\usepackage{graphicx,amsmath,bbm} 
\newcommand{\dir}{/home/stock/tex/Paper/Bib}

\date{\today}

\begin{document}

\author{Hiroshi Fujisaki$^1$}\email{fujisaki@theochem.uni-frankfurt.de} 
\author{Kiyoshi Yagi$^2$}\email{yagi@qcl.t.u-tokyo.ac.jp}
\author{Kimihiko Hirao$^2$}\email{hirao@qcl.t.u-tokyo.ac.jp}
\author{John E.\ Straub$^3$}\email{straub@bu.edu}
\author{Gerhard Stock$^1$}\email{stock@theochem.uni-frankfurt.de
} 
\affiliation{$^1$Institute of Physical and Theoretical Chemistry,
J.\ W.\ Goethe University, Max-von-Laue-Str.\ 7,                                    
60438 Frankfurt, Germany
}
\affiliation{$^2$Department of Applied Chemistry, 
School of Engineering, The University of Tokyo, 
Hongo 7-3-1, Bunkyo-ku, Tokyo 113-8656, Japan
}
\affiliation{$^3$Department of Chemistry,
Boston University, 590 Commonwealth Avenue,
Boston,  Massachusetts 02215, USA
}

\title{Quantum and classical vibrational relaxation dynamics of 
N-methylacetamide on ab initio potential energy surfaces}

\begin{abstract}
Employing extensive quantum-chemical calculations at the DFT/B3LYP and
MP2 level, a quartic force field of isolated N-methylacetamide
is constructed. Taking into account 24 vibrational degrees of freedom,
the model is employed to perform numerically exact vibrational
configuration interaction calculations of the vibrational energy
relaxation of the amide I mode. It is found that the energy
transfer pathways may sensitively depend on details of the theoretical
description. Moreover, the exact reference calculations were used to
study the applicability and accuracy of (i) the quasiclassical
trajectory method, (ii) time-dependent second-order perturbation
theory, and (iii) the instantaneous normal mode description of
frequency fluctuations. Based on the results, several
strategies to describe vibrational energy relaxation in biomolecular
systems are discussed.
\end{abstract}


\maketitle

%
%
\section{Introduction}
\baselineskip7.2mm

Amide modes in peptides and proteins are important probes to establish
a structure-spectroscopy relation for biomolecules.\cite{KB86,BZ02}
The amide I modes, which are localized at the C=O bond with
frequencies of 1600$-$1700 cm$^{-1}$, have attracted special attention
because they are sensitive to the secondary structure of
peptides.\cite{TT92} This work is concerned with the vibrational
energy relaxation of amide I modes. Recent experimental studies on
model peptides as well as various small globular peptides have shown
that the $\nu=1\rightarrow 0$ population decay time $T_1$ of the amide
I mode is about 1 ps for all systems
considered.\cite{HLH98,ZAH01,Tokmakoff06} To explain the ultrafast
energy and phase
relaxation  of amide I modes, 
a number of theoretical formulations have been
given,\cite{TT98,GCG02,Skinnergroup,HHC05,Mukamelgroup,BS06,Stock06,
NS03,KB07,DJBK07,Pouthier08,Leitner,FZS06,FYHS07} many of which focus on
N-methylacetamide (NMA, H$_3$C-COND-CH$_3$), a peptide-like small
molecule containing only a single amide I mode.  Employing standard
biomolecular potential energy functions (such as CHARMM\cite{CHARMM}
or GROMOS\cite{Gromacs}), quasiclassical trajectory
simulations\cite{NS03} as well as time-dependent perturbation theory
calculations\cite{FZS06} have been performed, qualitatively
reproducing the subpicosecond relaxation time found for NMA in
D$_2$O.\cite{HLH98,ZAH01,Tokmakoff06}

Alternatively, various groups considered
{\em ab initio} based potential energy functions to describe the
dynamics of amide I vibrations.
\cite{TT98,GCG02,Skinnergroup,HHC05,Mukamelgroup,BS06,Stock06,KB07,DJBK07,FYHS07}
Employing density functional theory (DFT) calculation at the
B3LYP/6-31G+(d) level, we recently constructed a quartic force field
of isolated NMA.\cite{FYHS07} The dynamics of the amide I vibration
was studied using the vibrational configuration interaction
method,\cite{bowman1,MULTIMODE,bowman2,FYHS07} including 24
vibrational degrees of freedom. The study indicated that the energy
transfer pathways may sensitively depend on details of the theoretical
description. Hence, one goal of the present paper is to examine the
sensitivity of amide I relaxation with respect to the {\em ab initio}
level (DFT/B3LYP and MP2), basis set (6-31G+(d) and 6-31G++(d,p)), and
description of the solvent (none or continuum model).

As the vibrational configuration interaction method gives a
numerically exact quantum-mechanical description of the vibrational
dynamics of the model system, it provides an ideal means to test
several approximations usually employed in the description of
vibrational energy relaxation.\cite{note1} The focus of this work is therefore to
study the applicability and accuracy of (i) the quasiclassical
trajectory method,\cite{NS03} (ii) (time-dependent) second-order
perturbation theory,\cite{Skinner,FZS06} and (iii) the instantaneous normal
mode description\cite{Stratt95} of frequency fluctuations. Based on
the results, we discuss several strategies to describe vibrational
energy relaxation in biomolecular systems.

\newpage
%
%
\section{Theory and methods}

\subsection{Quartic force field}
\label{sec:3MR}

While there are various ways to construct {\em ab initio} potential
energy surfaces (PES),\cite{GPNK06,YHH06} we employ here a simple but
accurate method called quartic force field, which is useful if we
consider vibrational dynamics around a single equilibrium state of a
molecule.\cite{rauhut,Christiansen,YHH07} In this approach, the PES is
approximated by a fourth-order Taylor expansion around the equilibrium
geometry
\begin{equation}
\tilde{V}(\{Q_i\})
=
\frac{1}{2}\sum_k \omega_k^2 Q_k^2 
+ \frac{1}{3!}\sum_{k,l,m} t_{klm} Q_kQ_lQ_m  
+ \frac{1}{4!}\sum_{k,l,m,n}u_{klmn} Q_k Q_l Q_m Q_n,
\label{eq:PES}
\end{equation}
where $Q_k$ and $\omega_k$ denote the $k$th normal coordinate and the
associated harmonic frequency, respectively, and the coefficients
$t_{klm}$ and $u_{klmn}$ represent the third- and fourth-order
derivatives of the PES.  The above expression can be recast in the
form of the $n$-mode coupling representation ($n$MR)
\cite{CCB97,YHTSG04,YHH07}
\begin{eqnarray}
V(\{Q_i\})
&=&
V^{1MR}(\{Q_i\})
+V^{2MR}(\{Q_i\})
+V^{3MR}(\{Q_i\}),
\\
V^{1MR}(\{Q_i\})
&=&
\sum_i 
\left( 
\frac{1}{2} \omega_i^2 Q_i^2 +
\frac{1}{6} t_{iii} Q_i^3 +
\frac{1}{24} u_{iiii} Q_i^4 
\right),
\\
V^{2MR}(\{Q_i\})
&=&
\sum_{i \neq j} 
\left( \frac{1}{2} t_{ijj} Q_i Q_j^2 + \frac{1}{6} 
u_{ijjj} Q_i Q_j^3 \right)
+\sum_{i < j} 
 \frac{1}{4} u_{iijj} Q_i^2 Q_j^2,
\\
V^{3MR}(\{Q_i\})
&=& 
\sum_{i<j<k} 
t_{ijk} Q_i Q_j Q_k 
+ 
\sum_{i \neq j<k} 
\frac{1}{2}u_{iijk} Q_i^2 Q_j Q_k. 
\end{eqnarray}
Note that we have omitted the 4MR terms $V^{4MR}(\{Q_i\}) =
\sum_{i<j<k<l} \frac{1}{4!} u_{ijkl} Q_i Q_j Q_k Q_l$.  This strategy
was successfully applied to several molecules.\cite{YHTSG04,Barone,
TYG05,KK07} By adding the normal mode kinetic energy $K=\sum_k
P_k^2/2$, we obtain the full approximate vibrational Hamiltonian
$H=K+V(\{ Q_i \})$ for the system.

In this study, the quartic force field was derived using the B3LYP or
MP2 methods with the 6-31G+(d) or 6-31G++(d,p) basis set. First, the
equilibrium structure and the harmonic frequencies were obtained for
trans-NMA using the Gaussian03 program
package.\cite{Gaussian03} Then, the third and fourth-order derivatives
were calculated by numerical differentiation of the analytic Hessian
\cite{YHH07} using the S{\footnotesize INDO} code developed by Yagi
\cite{sindo} or by the ``anharmonic'' option of Gaussian03.

%
%
\subsection{Vibrational configuration interaction method}

In a first step, we calculated the vibrational energies and
eigenstates of the above described model of NMA using the vibrational
configuration interaction (VCI) method developed by Bowman and
coworkers.\cite{bowman2} To generate basis functions for the VCI
calculations, we first performed a vibrational self-consistent field
(VSCF) calculation of the vibrational ground state. The VSCF
wave function is expressed as a direct product of one-mode functions or
{\it modals}\cite{Bowman86} as
\begin{eqnarray}
 \Phi_{\mathbf{n}}^{\mathrm{VSCF}} = \prod_{i=1}^{f} \phi_{n_i}^{(i)}(Q_i),
\end{eqnarray}
where {\bf n} and $f$ denote the vibrational quantum numbers and 
the number of degrees of freedom, respectively. The modals are
determined by the VSCF equation 
\begin{eqnarray}
 \left[ -\frac{1}{2}\frac{\partial^2}{\partial Q_i^2} + 
 \left\langle \prod_{j \neq i} \phi_{n_j}^{(j)} |
 V | \prod_{j \neq i} \phi_{n_j}^{(j)} \right\rangle \right] 
 \phi_{n_i}^{(i)} =
 \epsilon_{n_i}^{(i)} \phi_{n_i}^{(i)},
\end{eqnarray}
which yields the vibrational ground state ({\bf n=0}) as well as
the virtual modals. The VCI wave function is expressed as a 
linear combination of VSCF configurations
\begin{eqnarray}
 \Psi_{\mathbf{n}}^{\mathrm{VCI}} = \sum_{\mathbf{m}} C_{\mathbf{mn}} 
 \Phi_{\mathbf{m}}^{\mathrm{VSCF}}.
\end{eqnarray}
The VCI wave functions and energy levels are obtained by diagonalization of the 
VCI matrix
\begin{eqnarray}
 H_{\mathbf{mn}} = \langle \Phi_{\mathbf{m}}^{\mathrm{VSCF}} | H 
 | \Phi_{\mathbf{n}}^{\mathrm{VSCF}} \rangle.
\label{eq:VCImatrix}
\end{eqnarray}
All VSCF/VCI calculations were carried out using the S{\footnotesize
INDO} code \cite{sindo} for non-rotating molecules. The modals were
expanded in terms of harmonic oscillator wave functions. The number of
wave functions employed were 11, 9, 7, and 5 for \{$\phi^{(7)},
\phi^{(8)}$\}, \{$\phi^{(9)}$- $\phi^{(12)}$\},
\{$\phi^{(13)}$-$\phi^{(23)}$\}, and \{$\phi^{(24)}$-$\phi^{(30)}$\},
respectively. The mode index was labeled in increasing order of the
frequency and the six lowest-lying modes were kept frozen. The VSCF
configurations were selected by increasing energy until a cut-off
energy of $\sim 5000$ cm$^{-1}$, resulting in $N_{\rm CI} \simeq 6000$.

Once all eigenvalues \{$E_\mathbf{n}$\} and 
eigenfunctions \{$\Psi_\mathbf{n}^\mathrm{VCI}$\} are obtained, 
it is straightforward to calculate the time-dependent wave function
\begin{eqnarray}
|\Psi(t) \rangle =
\sum_\mathbf{n} \langle \Psi_\mathbf{n}^\mathrm{VCI} | \Psi(0) \rangle 
e^{-iE_\mathbf{n}t/\hbar} | \Psi_\mathbf{n}^\mathrm{VCI} \rangle,
\end{eqnarray}
which represents a numerically exact description of the above described
quartic force field model. 

In all calculations discussed below we assume that the initial state
is given by a single VSCF configuration $\Psi(0) =
\Phi_\mathbf{i}^\mathrm{VSCF}$, representing the $n=1$ state of the
amide I vibration. We thus obtain for the wave function 
\begin{equation}
|\Psi(t) \rangle
=
\sum_{\mathbf j}
O_{\mathbf j}(t) | \Phi_\mathbf{j}^\mathrm{VSCF} \rangle
\end{equation}
with 
 \begin{eqnarray}
O_\mathbf{j}(t) =
\sum_\mathbf{n} 
C_\mathbf{jn} C_\mathbf{ni}
e^{-iE_\mathbf{n} t/\hbar}. 
 \end{eqnarray}
and $P_\mathbf{j}(t) = |O_\mathbf{j}(t)|^2$ for the time-dependent
probability of state $\Phi_\mathbf{j}^\mathrm{VSCF}$. This allows us
to define the harmonic energy of vibrational mode $i$ as
\begin{equation}
E_i(t)= \hbar \omega_i \sum_{\mathbf j} P_{\mathbf j}(t) n^{(i)}_{\mathbf j},
\label{eq:qmode}
\end{equation}
where the zero point energy of the mode was disregarded. Equation
(\ref{eq:qmode}) will be used below to discuss the vibrational energy
relaxation in NMA.

%
%
\subsection{Time-dependent perturbation theory}
\label{sec:perturbation}

Employing time-dependent perturbation theory, recently Fujisaki {\em et
al.} have extended Fermi's Golden Rule to the non-Markovian
regime.\cite{FZS06,FS07} As a stringent test of this formulation, we
wish to compare the perturbative results to the exact VCI results
obtained for NMA. Assuming zero temperature, the perturbative
expression for the amide I ground state population is given
by\cite{FZS06} 
\begin{eqnarray}
P_0(t) &=& \rho_{00}(t) \simeq
\frac{\hbar}{8 {\omega}_S}
\sum_{\alpha,\beta} 
\frac{t_{S \alpha \beta}^2}{\omega_{\alpha} \omega_{\beta}}
\frac{1-\cos(\tilde{\omega}_S-\omega_{\alpha}-\omega_{\beta})t}
{(\tilde{\omega}_S-\omega_{\alpha}-\omega_{\beta})^2} 
\nonumber
\\
&\equiv& 
\sum_{\alpha,\beta} F_{S \alpha \beta}^2 
[1-\cos(\tilde{\omega}_S-\omega_{\alpha}-\omega_{\beta})t] ,
\end{eqnarray}
where ${\omega}_S$ is the is the anharmonicity-corrected system
frequency, $\omega_{\alpha}$ denote the (harmonic) frequency of the
bath modes, and $t_{S \alpha \beta}$ represents the third-order
coupling elements.\cite{FZS06}. Moreover, we have introduced a Fermi
resonance parameter\cite{Cremeens06}
\begin{equation}
F_{S \alpha \beta}
\equiv 
\left|
\sqrt{
\frac{\hbar}{2 {\omega}_S}
\frac{\hbar}{2 {\omega}_{\alpha}}
\frac{\hbar}{2 {\omega}_{\beta}}
}
\frac{t_{S \alpha \beta}}
{\hbar ({\omega}_S-\omega_{\alpha}-\omega_{\beta})}
\right|,
\label{eq:FRP}
\end{equation}
which is proportional to 
\begin{equation}
\eta \equiv
\left| 
\frac{\langle i| \Delta V|f \rangle}{\Delta E}
\right| ,
\end{equation}
where $|i \rangle$ and $|f \rangle$ are the initial and final {\it
harmonic} states, $\Delta V = V-\sum_k \omega_k^2 Q_k^2 /2$ denotes
the anharmonic vibrational coupling, and $\Delta E$ is the energy
difference between $|i \rangle$ and $|f \rangle$. The Fermi resonance
parameter is a key ingredient in the interpretation of vibrational
energy transfer processes.\cite{FZS06} This concept has been useful
for the classical dynamics characterization of the vibrational
relaxation pathways in water \cite{OT90} and in
myoglobin.\cite{MMK00}

%
%
\subsection{Classical description}
\label{sec:classical}

Quasiclassical trajectory calculations are a well-established approach
to approximately calculate the energy redistribution in molecular
systems.\cite{Schinkebook} In this method, the initial state of the
quantum system (e.g., the $n=1$ state prepared by an infrared laser
pulse) is represented by a phase-space distribution (e.g., the Wigner
distribution), which is sampled by an ensemble of classical
trajectories.  Starting by construction with the correct initial state
and recalling that classical mechanics can be considered as a
short-time approximation to quantum mechanics, a classical simulation
should be a good approximation to quantum-mechanics in the case of
ultrafast relaxation dynamics.\cite{Ankerhold02}

In this work, we solved Newton's equation on the PES defined in
Eq.~(\ref{eq:PES}), using Yoshida's sixth-order symplectic integrator
with a 0.5 fs time step.\cite{Yoshida90} To generate classical initial
conditions for the positions and momenta, we represent the normal
modes $\{Q_k,P_k\}$ in terms of classical action-angle variables
$\{n_k,\phi_k\}$ \cite{Schinkebook}
\begin{eqnarray}
Q_k &=& \sqrt{(2n_k+\gamma)\hbar /\omega_k}\sin\phi_k, 
\label{eq:ini1a} 
\\
P_k &=& \sqrt{(2n_k+\gamma) \hbar \omega_k}\cos\phi_k, 
\label{eq:ini1b} 
\end{eqnarray}
where the factor $\gamma = 1$ accounts for the zero-point energy of
the mode. To obtain the initial positions and momenta of the initially
excited amide I mode, we associate the action $n_k$ with the initial
quantum state of the amide I mode, i.e., $n_k=1$ for the first excited
state. The remaining modes are for $T=0$ K initially in the ground
state $n_k=0$ and only vibrate according to their zero-point
energy. In all cases, the vibrational phases $\phi_k$ are picked
randomly from the interval $[0,2\pi]$. Using this initial conditions,
the classical energy content of vibrational mode $i$ is defined as
\begin{eqnarray}
E_i(t)= \frac{\langle P_i^2(t) \rangle}{2}+ \frac{\omega_i^2}{2}
\langle Q_i^2(t)\rangle - E_i^{\rm ZPE} ,
\label{eq:cmode}
\end{eqnarray}
where $\langle \dots \rangle$ represents an ensemble average over 100 
nonequilibrium trajectories and $E_i^{\rm ZPE}$ denotes the zero-point
energy of mode $i$.

Although Eqs.\ (\ref{eq:ini1a}) and (\ref{eq:ini1b}) represents a
correct quasiclassical representation of the initial state of the
system (in the sense that the vibrational energy distribution is
correct), it may give rise to unphysical behavior due to the
well-known zero-point energy problem of classical
mechanics.\cite{Guo96} In quantum mechanics, each oscillator mode must
hold an amount of energy that is larger or equal to the zero-point
energy of this mode. In a classical trajectory calculation, on the
other hand, energy can flow among the modes without this
restriction. In the case of our {\em ab initio} model of NMA, for
example, we have found that the high zero-point energy contained in
the C-H stretch modes ($\approx$ 1600 cm$^{-1}$) may be transferred to
a large amount to low-frequency modes of NMA, thus yielding unphysical
and unstable trajectories.

Various approaches have been proposed to fix the zero-point energy
problem,\cite{Guo96} however, most of these techniques share the
problem that they manipulate {\em individual} trajectories, whereas
the conservation of zero-point energy should correspond to a virtue of
the {\em ensemble average} of trajectories. Here, we adopt the method
introduced in Ref.\ \onlinecite{Stock99}, which invokes quantum
corrections to the classical calculation in order to restrict the
classically accessible phase space according to the rules of quantum
mechanics. At the simplest level of the theory, these corrections have
been shown to correspond to including only a fraction $\gamma$ ($0\le
\gamma \le 1$) of the full zero-point energy into the classical
calculation. The quantum correction $\gamma$ can be determined by
requiring that some observables of the system are
reproduced,\cite{Stock99} e.g., that the amide I energy remains larger
than the zero-point energy for all times under consideration.  In the
present case, we assigned full zero-point energy to the initially
excited amide I mode, while for the remaining modes we choose $\gamma$
according to the classical thermal energy at $T =300$ K.  This
protocol is simple to implement in standard MD codes and was shown to
give good results in previous quasiclassical trajectory studies of
vibrational energy redistribution.\cite{NS03}

\newpage
%
%
\section{Results and discussion}
\label{sec:results}
\subsection{Quantum dynamics at various levels of {\em ab initio} theory}

\begin{figure}[htbp]
\hfill
\begin{center}
\includegraphics[scale=1.2]{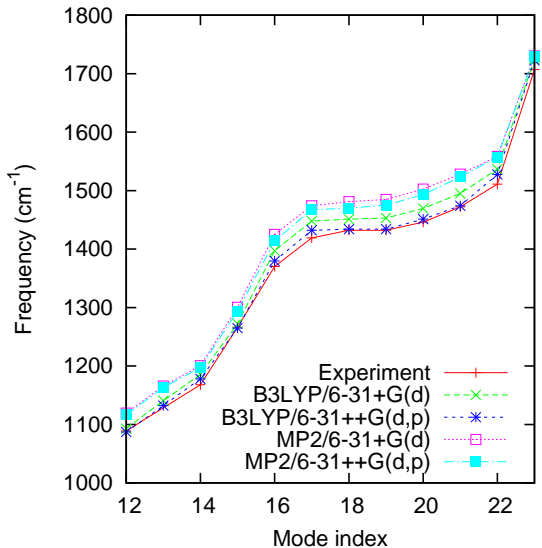}
\end{center}
\caption{Vibrational frequencies of isolated N-methylacetamide (NMA),
  calculated from the quartic force field using the VSCF/VCI method.
  Compared are experimental\cite{ATT84} and calculated frequencies at
  various levels of theory. The amide I vibration corresponds to mode
  \# 23 and amide II to \# 22.}
\label{fig:freq}
\end{figure}

The theoretical description of the amide I vibration of NMA first of
all depends on the quantum-chemical parameterization of the quartic
force field. Here, the following four levels of theory were employed:
DFT/B3LYP with 6-31+G(d) or 6-31++G(d,p) basis set and MP2 with
6-31+G(d) or 6-31++G(d,p) basis set. To facilitate the comparison to
experiment, we first consider the vibrational frequencies of NMA, as
obtained from the diagonalization of the VCI matrix
(\ref{eq:VCImatrix}). Figure \ref{fig:freq} compares these frequencies
to experimental values obtained under argon matrix
conditions,\cite{ATT84} focusing on vibrational modes between 1000 and
2000 cm$^{-1}$.  MP2 somewhat overestimates the anharmonic frequencies
in comparison to DFT/B3LYP. Applying the 6-31+G(d) basis set, we
obtain an overall deviation (mean absolute deviation) of 52 and 26
cm$^{-1}$ at the MP2 and DFT/B3LYP level respectively. As the
improvement due to the larger 6-31++G(d,p) basis is relatively small
(compared to the increased effort), in the following we only show
results using the 6-31+G(d) basis.  We confirmed that the energy
transfer pathways are nearly the same in both cases.

\begin{figure}[htbp]
\hfill
\begin{center}
\begin{minipage}{.42\linewidth}
\includegraphics[scale=1.2]{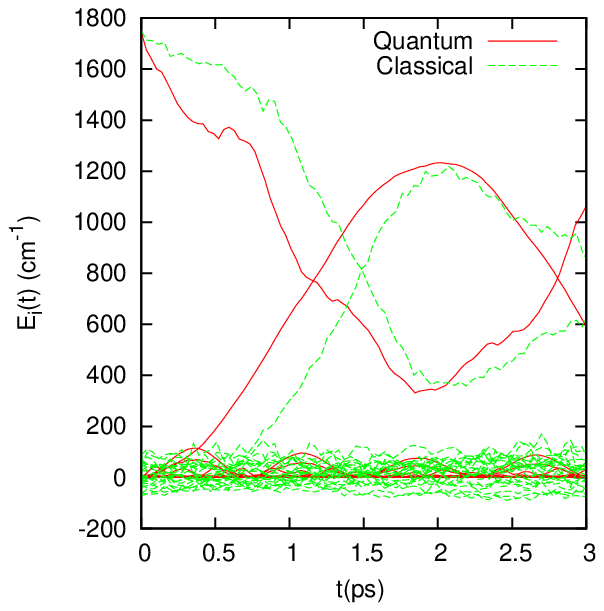}
\end{minipage}
\hspace{1cm}
\begin{minipage}{.42\linewidth}
\includegraphics[scale=1.2]{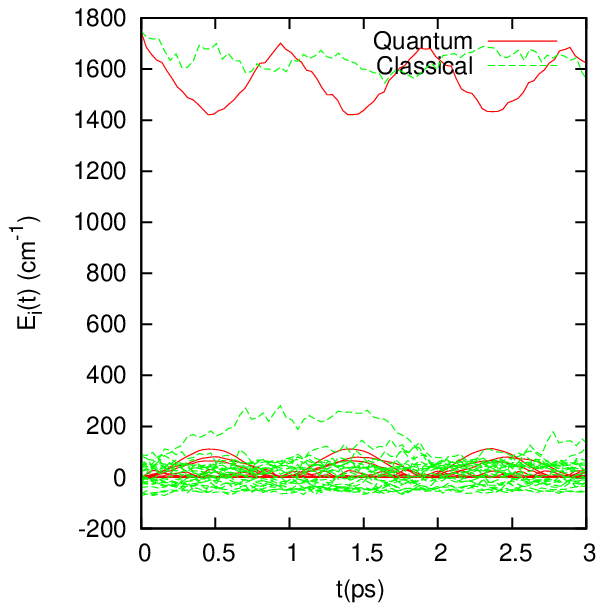}
\end{minipage}
\end{center}
\caption{Time evolution of the energy content of the
initially excited amide I mode as well as the remaining modes
of the NMA. Compared are quantum (red lines) and classical (green
lines) calculations, obtained at the DFT/B3LYP (left) and MP2 (right)
level of theory.}
\label{fig:vac1}
\end{figure}

\begin{figure}[htbp]
\hfill
\begin{center}
\begin{minipage}{.42\linewidth}
\includegraphics[scale=1.1]{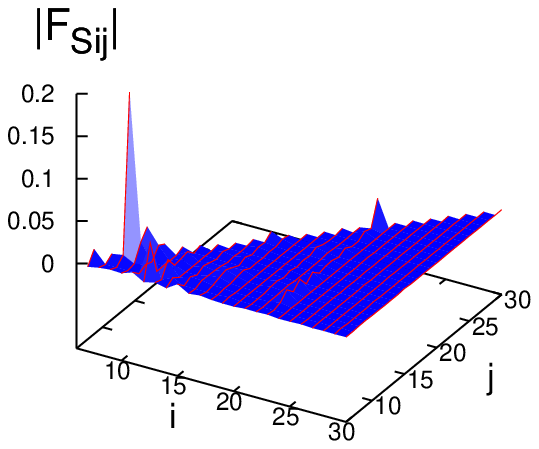}
\end{minipage}
\begin{minipage}{.42\linewidth}
\includegraphics[scale=1.1]{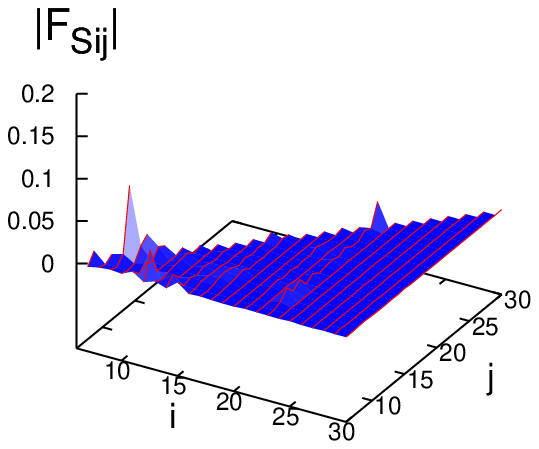}
\end{minipage}
\begin{minipage}{.42\linewidth}
\includegraphics[scale=1.1]{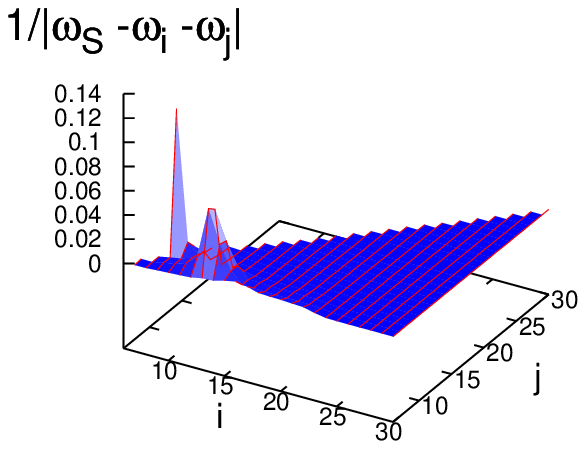}
\end{minipage}
\begin{minipage}{.42\linewidth}
\includegraphics[scale=1.1]{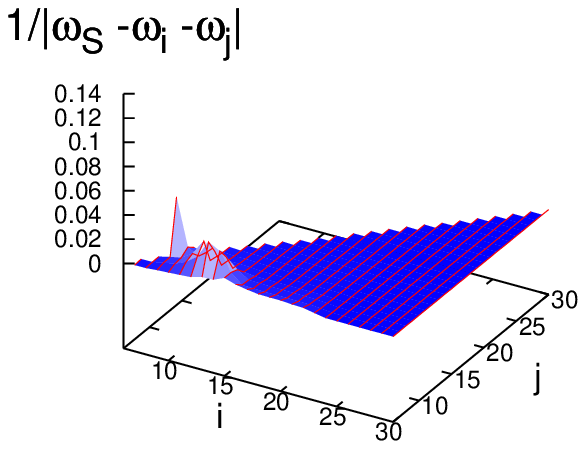}
\end{minipage}
\begin{minipage}{.42\linewidth}
\includegraphics[scale=1.1]{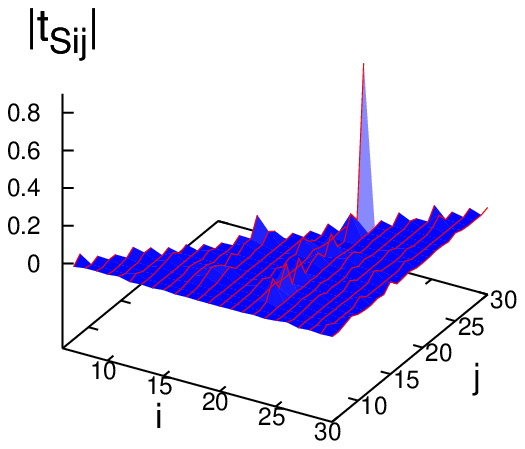}
\end{minipage}
\begin{minipage}{.42\linewidth}
\includegraphics[scale=1.1]{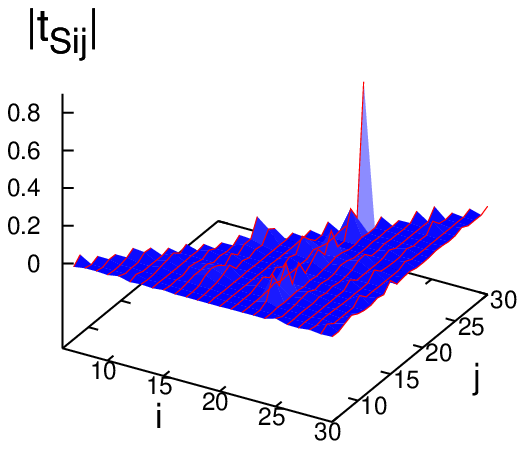}
\end{minipage}
\end{center}
\caption{Fermi resonance parameter $F_{S\alpha\beta}$ 
   [Eq.~(\ref{eq:FRP})] together with its resonance factor
  $|1/(\omega_S - \omega_\alpha - \omega_ \beta)|$ (cm) and the anharmonic
  coefficient $|t_{S\alpha\beta}|$ (kcal/mol/\AA).  Compared are quantum
  calculations, obtained at the DFT/B3LYP (left) and MP2 (right) level
  of theory.}
\label{fig:FRP}
\end{figure}

While all levels of theory at least qualitatively reproduce the
vibrational frequencies of NMA, the deviations may be more significant
when the vibrational energy relaxation of the amide I vibration is
considered. Figure \ref{fig:vac1} shows the time evolution of the
energy content [Eq.\ (\ref{eq:qmode})] of the initially excited amide
I mode (\# 23) as well as of the remaining modes of the molecule. As
discussed previously,\cite{FYHS07} in the B3LYP/6-31G+(d) calculation
most of the amide energy goes to the 9th vibrational mode at 869
cm$^{-1}$ via the pathway $|23_1\rangle \rightarrow |9_2\rangle$ and
to the 7th and 12th modes via the pathway $|23_1\rangle \rightarrow
|7_1 12_1\rangle$. As discussed in Ref.~\onlinecite{FYHS07}, these
normal modes correspond to motion of the C-terminal methyl-group atoms.  The
pathways can be explained by considering the Fermi resonance parameter
$F_{S\alpha\beta}$ defined in Eq.~(\ref{eq:FRP}), which describes
within time-dependent perturbation theory the intensity of transition
$|S_1\rangle \rightarrow |\alpha_1 \beta_1\rangle$.  Figure
\ref{fig:FRP} shows the Fermi resonance parameters for all mode
combinations $(\alpha, \beta)$ together with its resonance factor
$|1/(\omega_S - \omega_\alpha - \omega_ \beta)|$ and the anharmonic
coefficient $|t_{S\alpha\beta}|$.  Interestingly, it is the resonance
factor rather than the anharmonic coefficient that determines
$F_{S\alpha\beta}$ and thus the relaxation pathway. This underlines
the importance of an accurate calculation of vibrational frequencies.

Employing the force field obtained from the MP2/6-31G+(d)
calculations, Fig.\ \ref{fig:vac1} shows a qualitative change of the
amide I relaxation dynamics as compared to the results at the
DFT/B3LYP level. While the same vibrations are involved in the process
(i.e., mainly modes \# 9, and \# 7 and \# 12), the amide I energy
decreases only by $\approx$ 20 \% instead of $\approx$ 80 \%. Figure
\ref{fig:FRP} reveals that this is caused by the fact that the value
of the corresponding Fermi resonance parameters for MP2 are only half
of the value of the DFT parameters. (Note that $F_{S\alpha\beta}$
enters quadratically in the relaxation rate, which explains the
overall factor of four.)

\begin{figure}[htbp]
\hfill
\begin{center}
\begin{minipage}{.42\linewidth}
\includegraphics[scale=1.2]{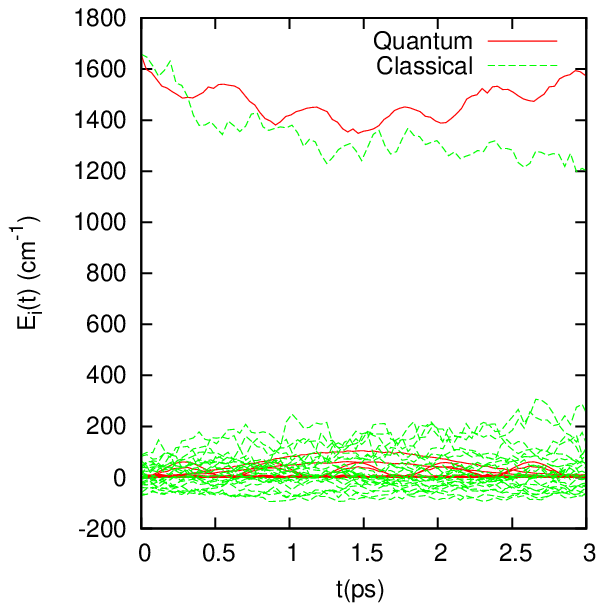}
\end{minipage}
\hspace{1cm}
\begin{minipage}{.42\linewidth}
\includegraphics[scale=1.2]{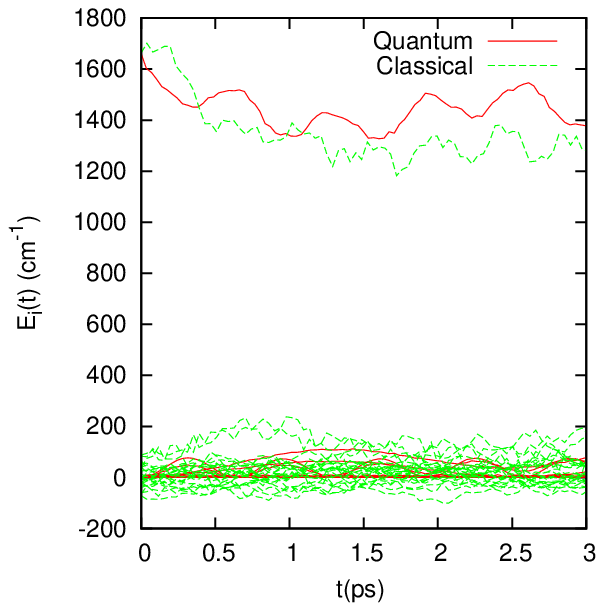}
\end{minipage}
\end{center}
\caption{Vibrational energy relaxation of NMA, obtained from DFT/B3LYP
calculations using an implicit solvent model.  Compared are quantum
(red lines) and classical (green lines) calculations, obtained for
implicit water (left) and implicit methanol (right).}
\label{fig:solv}
\end{figure}

As an approximative way to include solvent effects into the
parameterization of the quartic force field, we have employed the
implicit solvent model as implemented in Gaussian03\cite{Gaussian03}
and repeated the above calculations. As an example, Fig.\
\ref{fig:solv} shows results obtained for implicit water and implicit
methanol at the DFT/B3LYP level of theory. In both cases, one obtains
a relatively weak amide I relaxation, which mainly involves the modes
\#7, \#8, \#11, and \#12. 
However, it should be stressed that in experiment
\cite{HLH98,ZAH01} as well as in simulations using explicit water
solvent\cite{NS03,FZS06} there is a strong flow of vibrational energy
into the solvent degrees of freedom which is, of course, not captured
by an implicit model. Hence one should regard these calculations as
another example of energy transfer dynamics on different PESs.

%
%
\subsection{Classical relaxation dynamics}

The numerically exact quantum calculations shown in Figs.\
\ref{fig:vac1} and \ref{fig:solv} provide a stringent test for
quasiclassical trajectory simulations of vibrational energy transfer.
The comparison of both methods in these figures reveals that for all
cases considered the quasiclassical approximation yields at least
qualitative agreement with the reference calculation.  It is
interesting to note, though, that the classical results deviate
already at short times. This is a consequence of the fact that we
haven't employed the full zero point energy in initial conditions
(\ref{eq:ini1a}) and (\ref{eq:ini1b}). Doing so, we indeed obtain the
correct dynamics for the first few hundreds of femtoseconds, but also
observe unphysical flow of zero point energy at longer times (data not
shown).\cite{NS03} By employing full zero point energy in the initially
excited amide I mode and thermal energy in all remaining vibrational
modes, on the other hand, we obtain a reasonable classical
approximation.

Nonetheless, certain care should be taken when a quasiclassical
approximation to quantum dynamics is employed. In classical mechanics,
energy can flow among the vibrational modes without the restrictions
of quantum mechanics. For example, one may observe classical energy
transfer, if the initially excited vibrational mode $\omega_S$ is
about half of the frequency of a high-frequency bath mode
$\omega_\alpha$ (i.e., $ 2\omega_S \approx \omega_\alpha$), even
though the initially excited mode contains only the energy of a single
quantum. For the present force field, this behavior occurred when the
amide II mode was excited and energy transfer to the C-H stretch modes
was observed (data not shown). As in the case of the zero
point energy problem, it is therefore always advisable to check the
main energy pathways of a classical calculations, in order to make
sure that the classical energy flow could also happen in quantum
mechanics. 

%
%
\subsection{Perturbative description of relaxation dynamics}

\begin{figure}[htbp]
\hfill
\begin{center}
\begin{minipage}{.42\linewidth}
\includegraphics[scale=1.2]{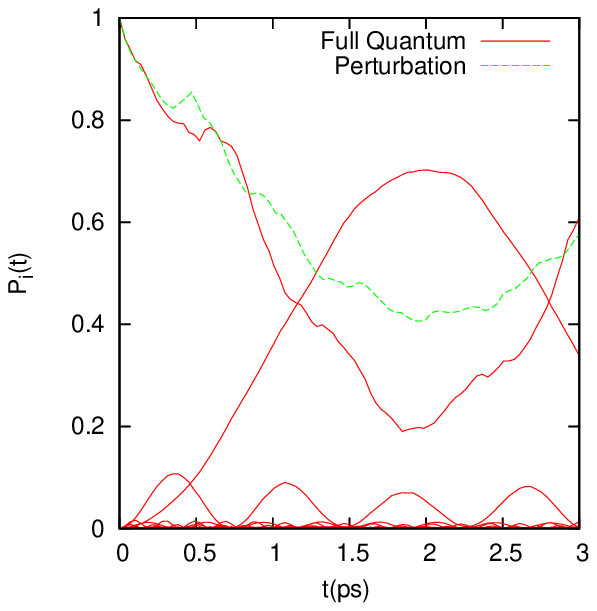}
\end{minipage}
\hspace{1cm}
\begin{minipage}{.42\linewidth}
\includegraphics[scale=1.2]{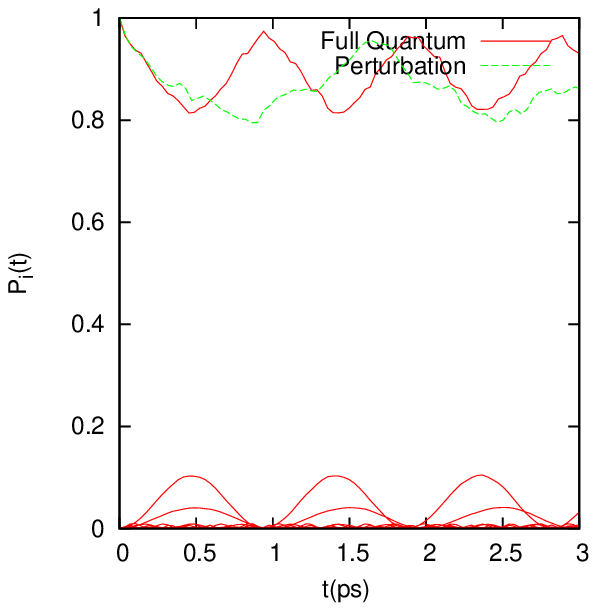}
\end{minipage}
\end{center}
\caption{Comparison of exact and perturbative calculation of
  vibrational energy relaxation of the amide I mode of NMA, employing
  (left) B3LYP and (right) MP2 level of theory.}
\label{fig:comp1}
\end{figure}

Apart from classical approximations, Golden Rule-type methods as
described in Sec.\ \ref{sec:perturbation} have be proven useful to
describe vibrational relaxation dynamics.\cite{FZS06,FS07} The
comparison of this method to the reference calculations in Fig.\
\ref{fig:comp1} reveals that the perturbative expression gives
quantitative results for the first few hundreds of femtoseconds. This
may be expected, because the initial process is given by the energy
transfer between the amide I mode and the doorway states (or first
tiers),\cite{Gruebele04} which is correctly taken into account by the
perturbative formula. At longer times, higher-order interactions and
the subsequent relaxation of the doorway states into other degrees of
freedom may become important, which is not included in the
second-order perturbative description. Nevertheless, in the present
case the perturbative calculation yields qualitatively correct results
for the first few picoseconds.

\begin{figure}[htbp]
\hfill
\begin{center}
\begin{minipage}{.42\linewidth}
\includegraphics[scale=1.2]{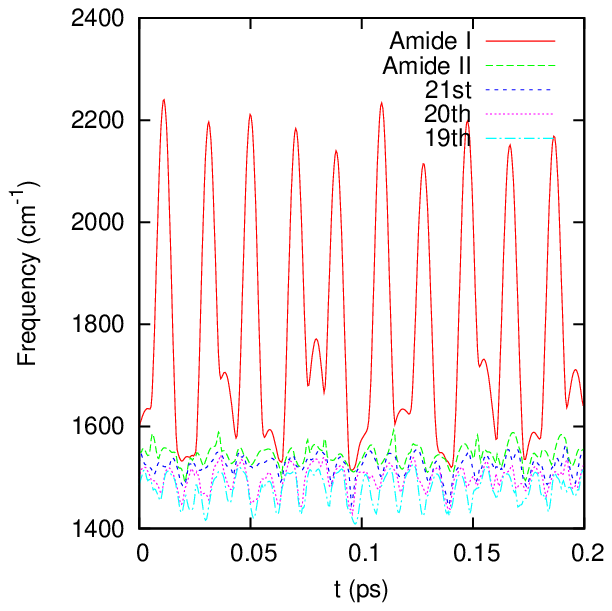}
\end{minipage}
\hspace{1cm}
\begin{minipage}{.42\linewidth}
\includegraphics[scale=1.2]{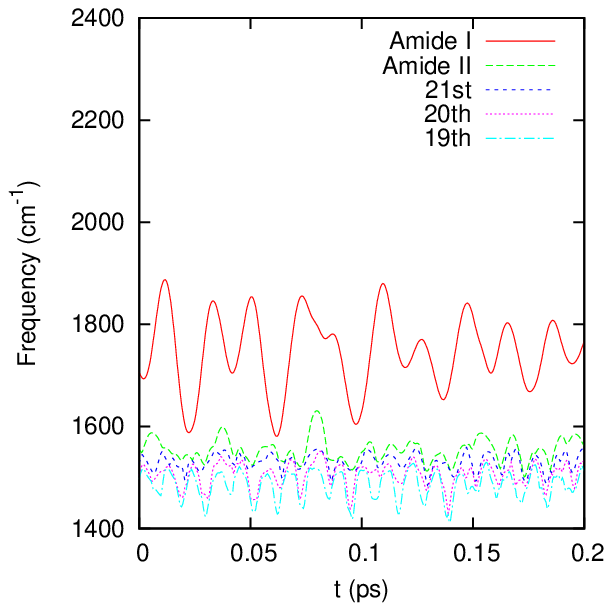}
\end{minipage}
\end{center}
\caption{Instantaneous normal mode frequencies of selected
  high-frequency modes of NMA, assuming (left) nonequilibrium and
  (right) equilibrium initial conditions, respectively.}
\label{fig:INM}
\end{figure}

A  possible approach to extend the application of Golden Rule-type
methods to the condensed phase is to take into account the
fluctuations of the solvent by calculating the instantaneous normal
mode frequencies of the solute.\cite{FS08} To study the validity of
this approach, we have calculated instantaneous normal mode
frequencies for the present model, assuming both equilibrium and
nonequilibrium initial conditions. As shown in Fig.~\ref{fig:INM}, the
instantaneous normal mode frequencies of the amide I mode and other
high-frequency modes of NMA may undergo fluctuations of several
hundred wavenumbers. Even under equilibrium conditions, the amide I
mode may vary by more than two hundred wavenumbers. This behavior is
quite unrealistic, when we recall that, in experiment, typically
fluctuations of tens of wavenumbers are observed.\cite{HLH98,ZAH01,Tokmakoff06} We
conclude that instantaneous normal mode frequencies cannot be regarded
as ``true'' vibrational frequencies, when strongly anharmonic {\em ab
initio} PES are employed.  A more realistic description of the
time-dependent frequencies is obtained, if we perform a geometry
optimization of the solute molecule, leading to ``quenched normal
modes''.\cite{OT90,FS08} Due to the absence of a fluctuating solvent,
however, in the present case this procedure only leads to the global
minimum of the molecule and therefore to time-independent (standard)
normal mode frequencies.

%
%
\section{Conclusions}
\label{sec:summary}

Following the previous work,\cite{FYHS07} we have constructed various
quartic force fields of isolated N-methylacetamide (NMA) and performed
numerically exact VSCF/VCI calculations of the vibrational energy
relaxation of the amide I mode of NMA. All levels of {\em ab initio}
theory considered (DFT/B3LYP and MP2 with 6-31+G(d) or 6-31++G(d,p)
basis set) reproduced the vibrational frequencies of NMA at least
qualitatively. However, the various theoretical levels were found to
lead to significant deviations of the vibrational energy
relaxation. This behavior could be explained through the Fermi
resonance parameter [Eq.\ (\ref{eq:FRP})], which includes the
anharmonic coupling as well as the resonance condition of the energy
transfer. The latter was found to predominantly determine the
efficiency of the process. It should be noted that this sensitivity to
the accurate description of vibrational frequency is particularly high
in the case of isolated NMA. Larger peptides and proteins in aqueous
solution provide a much higher vibrational level density, which
facilitates efficient relaxation, since there are always some
approximately resonant modes available.\cite{FS07} Combined with the overall
averaging effect of solvent-induced frequency fluctuations, this may
render a more qualitative quantum-chemical description sufficient.

Taking the VSCF/VCI results as reference calculations, we have studied
the validity and accuracy of various approximations usually employed.
It was found that second-oder perturbation theory provides a faithful
description of the first few hundreds of femtoseconds of the
relaxation dynamics. This is because the initial energy transfer
between the amide I mode and the doorway states is correctly described
by the perturbative formula. At longer times, this approximation
yields only qualitatively correct results, since higher-order
interactions and the subsequent relaxation of the doorway states
into other degrees of freedom may become important. To extend this
method to the condensed phase, instantaneous normal modes are often
invoked in order to account for the solvent-induced frequency
fluctuations of the system.  In the present case of strongly
anharmonic {\em ab initio} PES, however, the resulting instantaneous
normal mode frequencies were found to undergo unrealistically high
fluctuations of several hundred wavenumbers. A more realistic
description of the time-dependent frequencies may be obtained, if we
perform a geometry optimization of the solute molecule, leading to
``quenched normal modes''.\cite{OT90,FS08}

We have also performed quasiclassical trajectory simulations of
vibrational energy transfer, which were found to be in at least
qualitative agreement with the reference calculation. However, in
classical mechanics, energy can flow among the vibrational modes
without the restrictions of quantum mechanics. A well-known example is
the zero-point energy problem, which in the present case was
circumvented by employing full zero point energy only in the initially
excited amide I mode and thermal energy in all remaining vibrational
modes. Moreover, in classical mechanics energy transfer may also occur
to high-frequency bath modes. To assure that the classical energy flow
also would happen in quantum mechanics, it is therefore always
advisable to check the main energy pathways of a classical
calculation.

\acknowledgments 
We thank P.H. Nguyen, S.M. Park, and A. Dreuw for
discussions and helpful comments. HF is grateful to the Alexander von
Humboldt Foundation for their generous support. This work has been
supported by the Frankfurt Center for Scientific Computing and the
Fonds der Chemischen Industrie.

%
%

\bibliography{\dir/classic,\dir/femto,\dir/md,\dir/stock,new}

\bibliography{\dir/md,\dir/stock,new}

\end{document}